\documentclass[aps,prb,twocolumn,longbibliography,superscriptaddress,floatfix,10pt]{revtex4-2}

\usepackage[utf8]{inputenc}
\usepackage{amssymb,amsfonts,amsmath,dsfont}
\usepackage{graphicx}
\usepackage{textcomp}
\usepackage{gensymb}
\usepackage{pifont}
\usepackage{bm}
\usepackage{color}
\usepackage{accents}
\usepackage{mathtools}
\usepackage{nicefrac}
\usepackage{url}
\usepackage{multirow}
\usepackage{hyperref}
\usepackage[capitalize]{cleveref}
\usepackage{tabularx}
\usepackage{upgreek}
\usepackage{chemformula}
\hypersetup{colorlinks=true,citecolor=blue,urlcolor=blue,linkcolor=blue}

\newcommand{\hp}{Holstein-Primakoff }
\newcommand{\dm}{Dyson-Maleev }
\newcommand{\dmi}{Dzyaloshinskii–Moriya }

\newcommand{\kb}{ \mathbf{k}}
\newcommand{\qb}{ \mathbf{q}}
\newcommand{\qbp}{ \mathbf{q'}}

\newcommand{\op}[1]{ \hat{#1}}
\newcommand{\ad}[1]{ \op{\alpha}^\dag}
\newcommand{\bo}[1]{ \op{\beta}}
\newcommand{\bd}[1]{ \op{\beta}^\dag}

\newcommand{\ak}[1]{ \op{a}_{\kb_{#1}}}
\newcommand{\akd}[1]{ \op{a}_{\kb_{#1}}^{\dag}}
\newcommand{\bk}[1]{ \op{b}_{\kb_{#1}}}
\newcommand{\bkd}[1]{ \op{b}_{\kb_{#1}}^{\dag}}

\newcommand{\apk}[1]{ \op{\alpha}_{\kb_{#1}}}
\newcommand{\apkd}[1]{ \op{\alpha}_{\kb_{#1}}^\dag}
\newcommand{\btk}[1]{ \op{\beta}_{\kb_{#1}}}

\newcommand{\aq}[1]{ \op{\alpha}_{\mathbf{-Q}}}
\newcommand{\aqd}[1]{ \op{\alpha}_{\mathbf{-Q}}^\dag}
\newcommand{\bq}[1]{ \op{\beta}_{\mathbf{Q}}}
\newcommand{\bqd}[1]{ \op{\beta}_{\mathbf{Q}}^\dag}

\newcommand{\boga}{ \eta_1}
\newcommand{\bogb}{ \eta_2}
\newcommand{\bogc}{ \eta_3}
\newcommand{\uvec}[1]{ \hat{#1}}
\newcommand{\den}{ \updelta n}
\newcommand{\dep}{ \updelta \phi}
\newcommand{\ds}[1]{\left( \frac{\delta_{#1}}{N_c} \right)^2 }

\begin{document}
\title{Stability of chiral magnon condensates in collinear antiferromagnetic insulators}

\author{Therese Frostad}
\thanks{These authors contributed equally to this work.}
\affiliation{Center for Quantum Spintronic, Department of Physics, Norwegian University of Science and Technology, NO-7491 Trondheim, Norway}
\author{Anne Louise Kristoffersen}
\thanks{These authors contributed equally to this work.}
\affiliation{Center for Quantum Spintronic, Department of Physics, Norwegian University of Science and Technology, NO-7491 Trondheim, Norway}
\author{Verena Brehm}
\affiliation{Center for Quantum Spintronic, Department of Physics, Norwegian University of Science and Technology, NO-7491 Trondheim, Norway}
\author{Roberto E. Troncoso}
\affiliation{Departamento de Física, Facultad de Ciencias, Universidad de Tarapacá, Casilla 7-D, Arica, Chile}
\author{Arne Brataas}
\affiliation{Center for Quantum Spintronic, Department of Physics, Norwegian University of Science and Technology, NO-7491 Trondheim, Norway}
\author{Alireza Qaiumzadeh}
\affiliation{Center for Quantum Spintronic, Department of Physics, Norwegian University of Science and Technology, NO-7491 Trondheim, Norway}

\date{\today}

\begin{abstract}
Quasiequilibrium magnon Bose-Einstein condensates in ferromagnetic insulators have been a field of great interest, while condensation in antiferromagnetic systems has not yet been explored in detail. 
We analyze the stability of condensed chiral magnons in two antiferromagnetic insulators: a uniaxial easy-axis system and a biaxial system.
We show that two-component magnon condensation and inter-magnon interactions are essential to create metastable magnon condensation. 
The uniaxial system with a Rashba-type \dmi interaction supports two degenerate condensate populations at finite wave vectors. We find that the condensation state in this model is (meta)stable only when the distribution of condensed magnons between the two populations is symmetric. In addition, we demonstrate the emergence of a zero-sound-like Goldstone mode in antiferromagnetic systems that support two-component magnon condensation.
On the other hand, in the biaxial system without \dmi interaction, we predict that the magnon condensate cannot stabilize due to the breaking of the magnon degeneracy. Our results suggest that this instability is a general characteristic of single-component quasiequilibrium quasiparticle condensates.
\end{abstract}

\maketitle

\section{Introduction}

Bose-Einstein condensation (BEC) is a celebrated phenomenon in quantum physics describing a phase transition where bosonic particles form an equilibrium macroscopic state and collectively behave as a single quantum entity characterized by a coherent wave function
\cite{kapitza1938viscosity,allen1938flow,anderson1995observation,bradley1995evidence,davis1995bose,makinen2024magnon,borovik1984long,bunkov1992persistent,BEC1}.
This phase transition often requires temperatures close to absolute zero.
When the temperature decreases, the thermal wavelength of the particles increases, leading to a macroscopic occupation of the lowest quantum state.  
Although BEC was initially observed in conserved particles in the equilibrium state, the phenomenon has been extended to incorporate condensation processes of emerging nonconserved bosonic quasiparticles in various condensed matter systems \cite{Yukalov_2012}. Exploring \textit{quasiequilibrium} BEC states of nonconserved quasiparticles, such as excitons \cite{wang2019evidence}, phonons \cite{misochko2004transient}, and magnons \cite{Demokritov2006}, in a quasistationary state, opens new avenues for studying quantum phenomena in diverse physical systems, from superfluidity to novel states of matter with potential applications in quantum technologies.

In 2006, Demokritov {\it{et al.}} observed  magnon BEC (mBEC) in a ferromagnetic (FM) film of yttrium iron garnet (YIG) at room temperature \cite{Demokritov2006}. 
In their experimental setup, nonequilibrium magnons were excited under microwave parallel pumping, and the magnon condensate was investigated using Brillouin light scattering. This prompted a discussion of the formation, coherence, stabilization, and lifetime of mBEC states in FM systems 
\cite{demidov2007thermalization,dzyapko2007direct,demidov2008observation,demidov2008magnon,dzyapko2008monochromatic, PhysRevLett.100.257202,Troncoso2011, nowik2012spatially,Li2013,serga2014bose,Sun_2017,Mohseni_2020, Sukhachov2021,PhysRevB.106.024423,PhysRevLett.129.259901,frostad2024anisotropy,xu2024fr}.
Most experimental studies focus on mBEC in FM materials induced by parametric pumping. However, recent experiments also successfully generated mBEC electrically through the spin transfer torque mechanism \cite{divinskiy2021evidence} and thermally through a rapid cooling process \cite{schneider2020bose}. 

The magnon spectra of the thin-film YIG have double degenerate minima at finite momenta because of the interplay between exchange and dipolar interactions.
This results in a nonuniform condensate ground state in real space, which has been confirmed by experimental measurements \cite{nowik2012spatially}. 
The phase and magnon distribution between the two degenerate condensates, along with the stability of the condensate state, are influenced by nonlinear magnon interactions  
\cite{Li2013,salman2017microscopic,frostad2024anisotropy}.

The two-component mBEC in FM YIG, formed at two distinct wave vectors, is characterized by the magnon populations in each condensate state and their relative phase. These parameters have been proposed as the foundation for a classical analog of qubit logic \cite{mohseni2022classical,adrianov2014magnon}. 
Theoretical and experimental studies have also proposed the possibility of Josephson oscillations and persistent spin currents in mBEC systems
\cite{PhysRevB.104.144414,nakata2014josephson,Troncoso2014}.

Antiferromagnetic (AFM) systems have recently gained new interest in spintronics 
\cite{jungwirth2016antiferromagnetic,RevModPhys.90.015005,PhysRevLett.118.137201}. The absence of parasitic stray fields and the ability to operate in the THz regime make AFM systems promising candidates for next-generation high-speed, spintronic nanotechnologies. Unlike ferromagnets, antiferromagnets host two types of magnons that are chiral counterparts of each other \cite{rezende2019introduction,shiranzaei2022thermal}.
Nonequilibrium magnon excitations in AFM insulators have recently been realized via spin-transfer torques and laser excitations 
\cite{simoncig2017generation,tzschaschel2017ultrafast}.
The possibility of parallel pumping in orthorhombic AFM systems was explored both theoretically and experimentally decades ago \cite{Yamazaki1972ParallelP0}. However, its successful demonstration in conventional AFM systems remains elusive within the framework of modern AFM spintronics. 

Although magnon condensation in FM systems has been studied extensively both experimentally and theoretically, the counterpart phenomena in AFM materials have been less investigated.
To the best of our knowledge, there are only a few theoretical studies on the generation and stability analysis of quasiequilibrium mBEC states in AFM systems
\cite{Fjaerbu2017,PhysRevB.99.014405,bunkov2018magnon,10.1063/1.5096409}. 
It is worth mentioning that the quantum critical Bose gas of magnons at cryogenic temperatures has already been observed in \textit{quantum} AFM systems in which the quantum phase transition to the BEC state is controlled by a dc magnetic field \cite{giamarchi2008bose,matsumoto2024quantum,boundsate}. 

In compensated and collinear AFM systems, dipolar interactions are negligible and their magnon spectra often have a minimum at the center of the magnetic Brillouin zone, i.e., at the $\Gamma$ symmetry point with zero momentum.
Magnons in collinear AFM systems have two opposite chiral modes. Although these two magnon modes are degenerate in uniaxial easy-axis AFM systems, the degeneracy can be broken by applying an external magnetic field or in the presence of a hard-axis magnetic anisotropy 
\cite{rezende2019introduction,shiranzaei2022thermal,machado2017spin,RevModPhys.90.015005}. 
In these cases, the minima of both magnon modes remain at the $\Gamma$ symmetry point.
On the other hand, in systems with broken inversion symmetry, an antisymmetric \dmi (DM) interaction is allowed, which lifts the degeneracy of the two chiral magnon modes, causing them to split along the momentum axis \cite{kawano2019designing}.
As a result, the two magnon modes have two degenerate minima at finite wave vectors.
This momentum-dependent band splitting of AFM magnons has been measured in 
$\alpha$-Cu$_2$V$_2$O$_7$ by inelastic neutron scattering \cite{PhysRevLett.119.047201}.
These AFM systems may facilitate the presence of a nonzero wave vector double condensate state.

To generate a quasiequilibrium mBEC state with dynamic coherence, a necessary condition is to inject a critical number of nonequilibrium incoherent magnons into the system. Consequently, nonlinear magnon interactions facilitate thermalization \cite{Mohseni_2020, bunkov2018magnon}, causing a fraction of magnons to accumulate at the bottom of the magnon bands, thereby suddenly enhancing occupation at the band minima. Another necessary condition for achieving a quasiequilibrium mBEC state is the stability of the condensed magnons at the band minimum following thermalization.
In this paper, we explore the stability of condensed magnons by analytical calculations on two distinct Néel-ordered AFM systems: a uniaxial easy-axis AFM system with a Rashba-type DM interaction and a biaxial AFM system without DM interaction. These two classes of AFM systems serve as generic models for many realistic AFM materials studied in recent advancements in AFM spintronics.
In the first case, the magnon bands exhibit Rashba-type splitting along the momentum axis, while in the second case, the magnon bands are split along the energy axis, similar to Zeeman-type splitting. Furthermore, we demonstrate the existence of a zero-sound-like collective excitation in two-component  magnon condensation in AFM systems.

The rest of the paper is organized as follows.
Section \ref{sec:spin} introduces the spin model Hamiltonians of our AFM systems.
Section \ref{sec:boson} is a short presentation of the \hp and \dm transformations. 
Section \ref{sec:uniaxial} introduces the first AFM system we study, which is a uniaxial AFM model with DM interaction. 
The \textit{zero-sound spectrum} of this system is calculated in Sec. \ref{sec:zerosound}. 
Section \ref{sec:biaxial} introduces the second AFM system; a biaxial AFM without DM interaction. In this system, we study the condensate stability when the interband magnon interactions are suppressed by applying a strong magnetic field.
We compare the two AFM systems and present our concluding remarks in Sec. \ref{sec:summary}.

\section{Spin Model Hamiltonians} \label{sec:spin}
We consider a collinear two-sublattice AFM system, where the sublattice spins 
$\mathbf{S}_A$ and $\mathbf{S}_B$ align antiparallel to each other along the Néel vector, which is oriented in the $\uvec{z}$ direction. This system is described by the following extended quantum Heisenberg model:
\begin{equation}
\mathcal{H} = 
\mathcal{H}_\text{ex} + \mathcal{H}_\text{ani} + \mathcal{H}_\text{D} + \mathcal{H}_\text{Z}
.\label{eq:h1}
\end{equation}
The Heisenberg exchange interaction $\mathcal{H}_\text{ex}$ acts between nearest-neighbor (NN) spins at lattice sites $i$ and $j$ with an AFM exchange coupling $J>0$,
\begin{equation}
\mathcal{H}_\text{ex} =
J \sum_{\substack{i \in A,B \\ j \in NN }} \mathbf{S}_i \cdot \mathbf{S}_j.
\end{equation}

The magnetic anisotropy Hamiltonian incorporates single-ion easy and hard-axes magnetic anisotropies \cite{shiranzaei2022thermal,rezende2019introduction},
\begin{align}
\mathcal{H}_\text{ani} =
- K_z \sum_{i} (S_i^z)^2
+ K_x \sum_{i} (S_i^x)^2.
\label{eq:h1ax}
\end{align}
Here, $K_z \geq 0$ parametrizes the easy-axis magnetic anisotropy along the 
$\uvec{z}$ direction, while $K_x \geq 0 $ denotes the hard-axis magnetic anisotropy along the $\uvec{x}$ direction. 

The Rashba-type DM Hamiltonian reads 
\begin{equation}
\mathcal{H}_\text{D} =
 \sum_{\substack{i \in A,B \\ j \in NN }} \mathbf{D}_{ij} \cdot( \mathbf{S}_i \times \mathbf{S}_j),
\end{equation}
where the DM vector is antisymmetric, $\mathbf{D}_{ij} = -\mathbf{D}_{ji}$. 
The vector is oriented along the $\uvec{z}$ direction so that 
$\mathbf{D}_{ij} = D \nu_{ij} \uvec{z}$, with $\nu_{ij} = \pm 1$ for clockwise and counterclockwise hopping directions.

Finally, the interaction between an external magnetic field along the easy-axis $\uvec{z}$ direction and localized spins is modeled by the Zeeman coupling, 
\begin{equation}
\mathcal{H}_\text{Z} = -\mu_\text{B} h \sum_i S_i^z,
\end{equation}
where $\mu_\text{B}$ is the Bohr magneton and $h$ is the magnetic field amplitude.

\section{Nonlinear boson representations in the theory of antiferromagnets} \label{sec:boson}
The elementary excitations of an ordered magnet are magnons. To describe these excitations, it is suitable to pass from spin representation to canonical boson operators. There are several bosonic representations of spins 
\cite{katanin2007magnetic,auerbach2012interacting}. In this paper, we consider both the \hp and \dm transformations, and we compare the two methods in Appendix \ref{sec:appendix}. 
In particular, the \dm transformations use spin raising and lowering operators that are not Hermitian conjugates of each other. For this reason, we also investigate whether the choice of bosonization technique affects results in Appendix \ref{sec:appendix}. Within our approximations, we could not find any physical difference between these two transformations.

\emph{I.} \hp transformation. In this transformation, we define raising and lowering spin operators which are Hermitian conjugates of each other, 
$\op{S}^+ = (\op{S}^-)^\dag = \op{S}^x + i \op{S}^y$.
We denote bosonic creation (annihilation) by 
$\op{a}_i^\dag $ ($\op{a}_i$) for spins on sublattice $A$ and 
$\op{b}_j^\dag $ ($\op{b}_j$) for spins on sublattice $B$. 
Given the quantization axis along the $\uvec{z}$ axis, we have the following transformation for spin operators at sublattice $A$,
\begin{subequations}
\begin{align}
\op{S}_i^+ &= \sqrt{2S - \op{a}_i^\dag \op{a}_i } \op{a}_i 
\approx \sqrt{2S } \biggl( \op{a}_i - \frac{\op{a}_i^\dag \op{a}_i \op{a}_i}{4S} \biggr)
, \\
\op{S}_i^z &= S - \op{a}_i^\dag \op{a}_i,
\end{align} \label{eq:hpa}
\end{subequations}
and sublattice $B$, 
\begin{subequations}
\begin{align}
\op{S}_j^+ &=  \op{b}_j^\dag \sqrt{2S - \op{b}_j^\dag \op{b}_j } 
\approx \sqrt{2S } \biggl( \op{b}_j^\dag - \frac{\op{b}_j^\dag \op{b}_j^\dag \op{b}_j }{4S} \biggr)
,\\
\op{S}_j^z &= - (S - \op{b}_j^\dag \op{b}_j).
\end{align} \label{eq:hpb}
\end{subequations}
To capture the leading nonlinear interactions in the spin Hamiltonian, we performed a Taylor expansion of the square roots in the above expressions up to third order in the bosonic operators, assuming a large spin length \( S \) and a low boson occupancy at each site. According to Ref. \cite{bunkov2018magnon}, the critical magnon number density, needed to generate mBEC state, may satisfy this approximation.

\emph{II.} \dm transformation.
In this representation, the raising and lowering spin operators are not Hermitian conjugates of each other $\op{S}^+ \neq (\op{S}^-)^\dag $.
The spin operators on sublattice $A$ are given by,
\begin{subequations}
\begin{align}
\op{S}_i^+ &= \sqrt{2S} \biggl(\op{a}_i - \frac{\op{a}_i^\dag \op{a}_i \op{a}_i}{2S} \biggr)
,\\
\op{S}_i^- &= \sqrt{2S} \op{a}_i^\dag 
,\\
\op{S}_i^z &= S - \op{a}_i^\dag \op{a}_i
,
\end{align} 
\end{subequations}
and on sublattice $B$ become, 
\begin{subequations}
\begin{align}
\op{S}_j^+ &= \sqrt{2S} \biggl(\op{b}_j^\dag -  \frac{\op{b}_j^\dag \op{b}_j^\dag \op{b}_j }{2S} \biggr),\\
\op{S}_j^- &= \sqrt{2S} \op{b}_j,\\
\op{S}_j^z &= - (S - \op{b}_j^\dag \op{b}_j).
\end{align} 
\end{subequations}
In both transformations, an additional constraint ensures that the boson occupancy at each lattice site does not exceed \(2S\), i.e., $\op{a}_i^\dag \op{a}_i (\op{b}_j^\dag \op{b}_j) \leq 2S$. This condition preserves the correct representation of the \(2S+1\) eigenstates of \(\hat{S}_i^z\).

We find that both bosonization transformations in our system lead to the same result. Therefore, in the following sections, we focus exclusively on applying the \hp transformation to the spin Hamiltonian (\ref{eq:h1}). A discussion of the \dm transformation is provided in Appendix \ref{sec:appendix}.

\section{Magnon BEC in uniaxial easy-axis AFM case} \label{sec:uniaxial}
First, we study a uniaxial AFM with DM interaction. In this case, we set $K_x=0$ in the anisotropy Hamiltonian, Eq. (\ref{eq:h1ax}).
We use the \hp transformation and the Fourier transform
$\op{a}_i = (N/2)^{-1/2} \sum \limits_{\kb} \ak{} \text{e}^{-i\kb\cdot {\bm{r}}_i}$, where $N$ denotes the number of spin sites, to express the total Hamiltonian in terms of canonical bosonic operators in momentum space.

\begin{figure}
\centering
\includegraphics[width=0.4\textwidth]{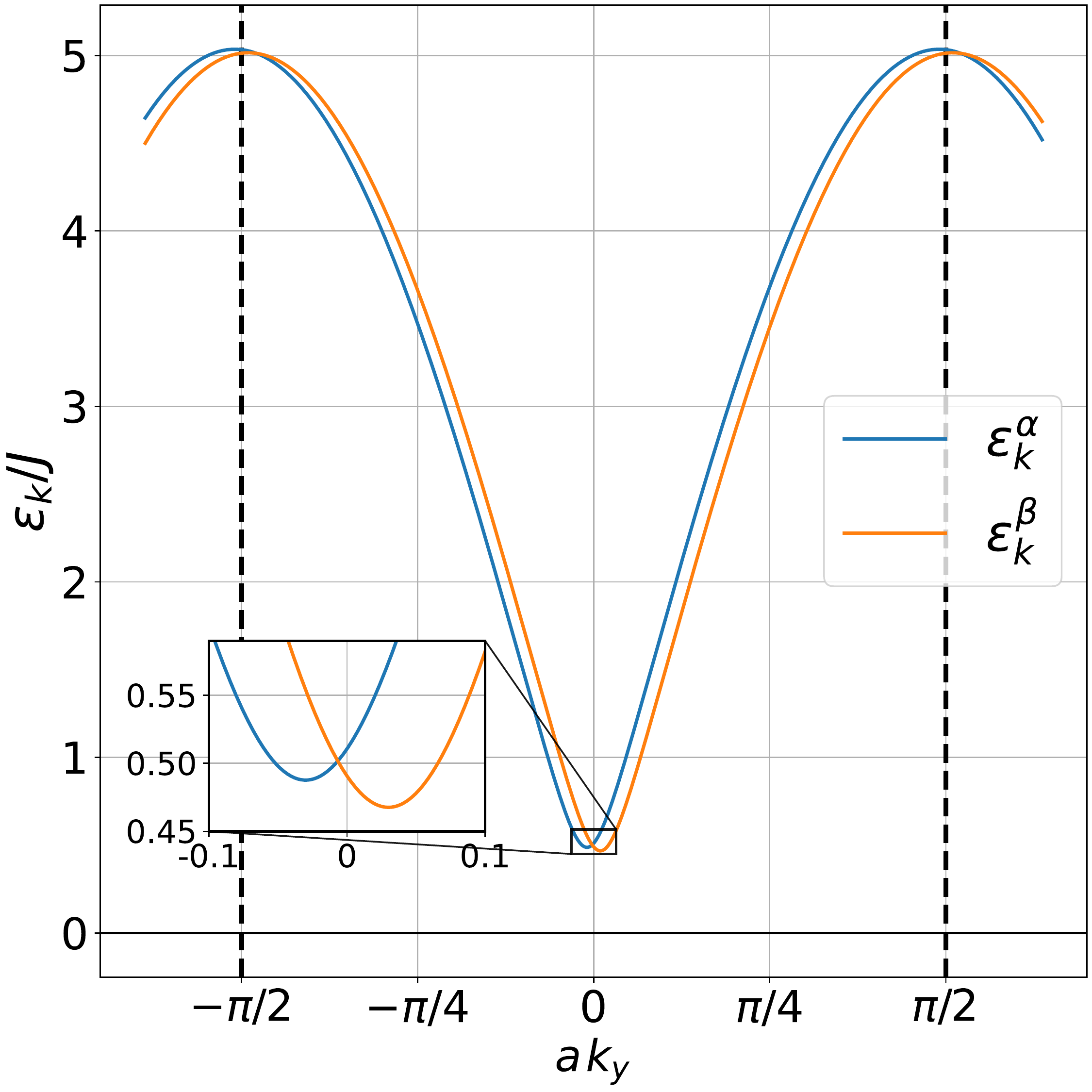}
\caption{Dispersion relations for the right-handed $\alpha$ (blue) and left-handed  $\beta$ (orange) magnons. We set 
$\omega_D / \omega_\text{ex} = 0.03$, 
$\omega_z / \omega_\text{ex} = 0.025$, 
$\omega_H S/ \omega_\text{ex} = 0.0025$, and $S=5/2$. 
The black dashed vertical lines mark the boundaries of the first magnetic Brillouin zone.}
\label{fig:disprel}
\end{figure}

The noninteracting Hamiltonian, composed of quadratic bosonic operators, is expressed as 
\begin{align}
\mathcal{H}^{(2)} 
= & SJ \sum\limits_{\kb}  \big(z [\akd{} \ak{} + \bkd{} \bk{} ] \nonumber \\ & \qquad \quad + 2\cos(a k_y) [\ak{} \op{b}_{-\kb} + \akd{} \op{b}_{-\kb}^\dag ] \big) \nonumber \\
&+ 2SK_z \,\sum\limits_{\kb} \bigl[ \akd{} \ak{} + \bkd{} \bk{} \bigr] \nonumber \\
&-4SD \sum\limits_{\kb} \sin(a k_y) \bigl[ \akd{} \op{b}_{-\kb}^\dag + \ak{} \op{b}_{-\kb} \bigr]\nonumber \\
& - \mu_\text{B} h \sum\limits_{\kb} \bigl[ \bkd{} \bk{} - \akd{} \ak{} \bigr],
\label{eq:h2}
\end{align}
where $z$ and $a$ denote the coordination number and lattice constant, respectively, and $\kb$ is the magnon wave vector. To diagonalize this bosonic Hamiltonian and transition to the magnon representation, we apply a Bogoliubov transformation, 
\begin{subequations}
\begin{align}
\apk{} &= u_{\kb} \ak{} + v_{\kb} \op{b}_{-\kb}^\dag 
, \\
\btk{} &= v_{-\kb} \op{a}_{-\kb}^\dag + u_{-\kb} \bk{},
\end{align} \label{eq:bog}
\end{subequations}
where bosonic $\apk{}$ and $\btk{}$ operators denote two species of magnons in the AFM system. 
The Bogoliubov coefficients are then expressed as
\begin{subequations}
\begin{align} 
u_{\kb}^2 &= \frac{\omega_{ez}}{2}
\left[\omega_{ez}^2 - \left(\omega_\text{ex}\cos(a k_y) - \omega_D\sin(a k_y)\right)^2\right]^{-\frac{1}{2}} + \frac{1}{2}, \label{eq:bogUniax1} 
\\
v_{\kb}^2 &= \frac{\omega_{ez}}{2}
\left[\omega_{ez}^2 - \left(\omega_\text{ex}\cos(a k_y) - \omega_D\sin(a k_y)\right)^2\right]^{-\frac{1}{2}} - \frac{1}{2},
\label{eq:bogUniax2}
\end{align} 
\end{subequations}
where we have introduced the following parameters: 
$\omega_\text{ex} = 2SJ$,
$\omega_z = SK_z$,
$\omega_{ez} = \omega_\text{ex} + \omega_z $,
$\omega_D = 2SD $, and
$\omega_H = \mu_\text{B} h/2 $.
Finally, the noninteracting Hamiltonian in the diagonalized magnon basis reads 
\begin{equation}
\mathcal{H}^{(2)} 
= \sum\limits_{\kb} \left[ 
\varepsilon_{\kb}^{\alpha} \apkd{} \apk{} + 
\varepsilon_{\kb}^{\beta} \op{\beta}_{-\kb} \op{\beta}_{-\kb}^\dag \right].
\label{eq:h2ab}
\end{equation}
The dispersion relations for the left-handed $\apk{}$ and right-handed $\btk{}$ chiral AFM magnons, propagating along the $y$ direction, are given by 
\begin{subequations}\label{energy}
\begin{align} 
\varepsilon_{\kb}^\alpha &=
\sqrt{\omega_{ez}^2 - \left[\omega_\text{ex} \cos(ak_y) - \omega_D \sin(ak_y)\right]^2} + \omega_H 
, \\
\varepsilon_{\kb}^\beta &=
\sqrt{\omega_{ez}^2 - \left[\omega_\text{ex} \cos(ak_y) + \omega_D \sin(ak_y)\right]^2} - \omega_H. 
\end{align} 
\end{subequations}
From these dispersion relations, it is evident that the DM interaction term leads to a momentum-dependent splitting of magnon bands, while a Zeeman field leads to an energy-dependent splitting of bands; see Fig. \ref{fig:disprel}.
In the exchange dominant regime, $\omega_\text{ex} > \omega_D$, the energy-degenerate minima of the two magnon bands occur at finite wave vectors $Q_y=\pm a^{-1} \tan^{-1}(\omega_D/\omega_\text{ex})$.

The lowest-order interaction Hamiltonian, which is quartic in boson operators, is given by
\begin{align}
\mathcal{H}^{(4)} = 
- &\sum_{\substack{\kb \qb\qbp}} \frac{f_J(\kb)}{(N/2)}
\biggl[
\ak{} \op{b}_{\kb + \qb + \qbp}^\dag \op{b}_{\qb} \op{b}_{\qbp} + \text{H.c.}
\nonumber \\
& ~ \, ~ \, ~ \, ~ \, ~ \, ~ \, ~ \, ~ \, 
+ \op{a}_{\kb + \qb + \qbp}^\dag \op{a}_{\qb} \op{a}_{\qbp} \bk{} + \text{H.c.}
\nonumber \\
& ~ \, ~ \, ~ \, ~ \, ~ \, ~ \, ~ \, ~ \, 
+ 4 \op{a}_{\qb - \kb}^\dag \op{a}_{\qb} \op{b}_{\qbp + \kb}^\dag \op{b}_{\qbp} 
\biggr]
\nonumber \\
- & \sum\limits_{\kb  \qb \qbp} \frac{K_z}{(N/2)}
\biggl[
\op{a}_{\qb - \kb}^\dag \op{a}_{\qb} \op{a}_{\qbp + \kb} ^\dag \op{a}_{\qbp}
+ \op{b}_{\qb - \kb}^\dag \op{b}_{\qb} \op{b}_{\qbp + \kb} ^\dag \op{b}_{\qbp}
\biggr]
\nonumber \\
+&\sum\limits_{\kb  \qb \qbp} \frac{f_D(\kb)}{(N/2)} \biggl[
\akd{} \op{b}_{\qb}^\dag \op{b}_{\qbp}^\dag \op{b}_{\kb + \qb + \qbp}
+ \ak{} \op{b}_{\kb + \qb + \qbp}^\dag \op{b}_{\qb} \op{b}_{\qbp} 
\nonumber \\
& ~ \, ~ \, ~ \, ~ \, ~ \, ~ \, ~ \, ~ \, 
-\op{a}_{\qb}^\dag \op{a}_{\qbp}^\dag \op{a}_{\qb + \qbp + \kb} \bkd{} 
-\op{a}_{\qb + \qbp + \kb}^\dag \op{a}_{\qb} \op{a}_{\qbp} \bk{} 
\biggr]
\label{eq:h4}
\end{align}
where we defined 
$f_J(\kb) = J\cos(a k_y)$ and $f_D(\kb) = D\sin(a k_y)$. Note that the external magnetic field, applied along the magnetic ground state, does not appear in the interaction Hamiltonian.

We now rewrite the nonlinear bosonic Hamiltonian of \cref{eq:h4} in terms of magnon operators, using the Bogoliubov transformations, Eq. (\ref{eq:bog}). 
To investigate mBEC stability, we are only interested in magnons at the bottom of each respective magnon band, i.e.,
$\op{\alpha}_{k=-Q}$ and $\op{\beta}_{k=+Q}$. 
Therefore, the nonlinear Hamiltonian for condensed magnons can be rewritten as the sum of intra- and inter-band interaction terms
$\mathcal{H}_Q^\text{(4)} = \mathcal{H}_Q^\text{intra} + \mathcal{H}_Q^\text{inter} $, where the intra- and inter-band terms read,  
\begin{subequations}
\begin{align}
\mathcal{H}_Q^\text{intra } &=
\mathcal{A}_Q \left( \aq{} \aq{} \aqd{} \aqd{} + \bq{} \bq{} \bqd{} \bqd{}  \right)
\label{eq:h4interact_a}
,\\
\mathcal{H}_Q^\text{inter } &=
\mathcal{B}_Q \left( \aq{} \aqd{} \bq{} \bqd{}   \right) \nonumber \\
&+\mathcal{C}_Q \big( \aq{} \bq{}  \bq{}  \bqd{} +\aq{} \aqd{} \aqd{} \bqd{}  \nonumber \\
& ~ \, ~ \, + \aq{} \aq{} \aqd{} \bq{} + \aqd{} \bq{} \bqd{} \bqd{}  \bigr) \nonumber \\
&+\mathcal{D}_Q \left( \aq{} \aq{} \bq{} \bq{} + \aqd{} \aqd{} \bqd{} \bqd{}  \right) 
. 
\label{eq:h4interact_b}
\end{align}
\end{subequations}
The amplitudes of different interactions between condensate magnons are given by 
\begin{subequations}
\begin{align}
\mathcal{A}_Q &= -\frac{2}{N} \large[
2\left(f_J(Q) + f_D(Q)\right) \bogc +2J \bogb \nonumber \\ & \qquad \qquad + K_z (\boga - 2\bogb) \large] 
,\\
\mathcal{B}_Q &= -\frac{2}{N} \large[
8(f_J(Q) + f_D(Q)) \bogc +4J (\boga - \bogb) \nonumber \\ & \qquad \qquad + 4K_z \bogb \large] 
,\\
\mathcal{C}_Q &= -\frac{2}{N} \large[
(f_J(Q) + f_D(Q)) (\boga + \bogb) +4J \bogc  \nonumber \\ & \qquad \qquad + 2K_z \bogc \large] 
,\\
\mathcal{D}_Q &= -\frac{2}{N} \left[
2(f_J(Q) + f_D(Q)) \bogc +2J \bogb + K_z \bogb \right],
\end{align}
\label{eq:prefacs}
\end{subequations}
where we defined the following coefficients: 
\begin{subequations}
\begin{align}
\boga &= u_{-Q}^4 + v_{-Q}^4 + 4 u_{-Q}^2 v_{-Q}^2 ,\\
\bogb &= 2 u_{-Q}^2 v_{-Q}^2 ,\\
\bogc &= -u_{-Q} v_{-Q} ( u_{-Q}^2 + v_{-Q}^2 ).
\end{align}
\end{subequations}
Our goal is to analyze the nonlinear potential energy of the emergent condensate magnons at the bottom of two magnon bands. 
We represent the two-condensate populations as macroscopic wave functions using Madelung's transformation, which expresses a bosonic operator in terms of a macroscopic condensate density and its associated phase \cite{Li2013}, 
\begin{subequations}
\begin{align}
\aq{}  & \rightarrow \sqrt{N_\alpha} \, \text{e}^{i\phi_\alpha}, \\
 \bq{} & \rightarrow \sqrt{N_\beta} \, \text{e}^{i\phi_\beta},
\end{align}
\end{subequations}
where $N_{\alpha (\beta)}$ is the number of condensed chiral $\alpha (\beta)$-magnons at the bottom of the bands, and $\phi_{\alpha (\beta)}$ refers to their macroscopic phase.
We define the total number of condensed magnons as $N_c = N_\alpha + N_\beta$, the phase sum as $\Phi = \phi_\alpha + \phi_\beta$, and the difference in the population of chiral condensate magnons as $\delta = N_\alpha - N_\beta$.
Finally, we find the following nonlinear potential energy for condensate magnons: 
\begin{subequations}
\begin{align}
\mathcal{V}^{(4)} &= \mathcal{V}_Q^\text{intra} + \mathcal{V}_Q^\text{inter} 
, \label{eq:v4} \\
\mathcal{V}_Q^\text{intra} &= \frac{N_c^2}{2} \mathcal{A}_Q
\biggl[ 1 + \biggl(\frac{\delta }{N_c}\biggr)^2 \biggr]
,\\
\mathcal{V}_Q^\text{inter} &=  \frac{N_c^2}{2} \biggl[
2 \mathcal{C}_Q \sqrt{1-\biggl(\frac{\delta}{N_c}\biggr)^2} \, \cos(\Phi) \nonumber \\
& \hspace{8mm} - \biggl(\mathcal{D}_Q \cos(2\Phi) + \frac{\mathcal{B}_Q }{2}\biggr)\biggl(\frac{\delta}{N_c}\biggr)^2 \nonumber \\
& \hspace{8mm} + \frac{\mathcal{B}_Q}{2} + \mathcal{D}_Q \cos(2\Phi) \biggr].
\end{align}
\label{eq:v4u}
\end{subequations}
We consider the total number of condensate magnons $N_c$ to be constant. In general, the number of bosonic quasiparticles is not conserved due to, e.g., Gilbert damping, so this assumption is only valid in a time-scale proportional to the magnon condensate lifetime.
We note that the interaction Hamiltonian of AFM magnons, \cref{eq:h4interact_a,eq:h4interact_b}, in its generalized form is directly comparable to the interaction Hamiltonian of FM magnons with dipolar interactions \cite{frostad2024anisotropy}. 

To determine whether the generated condensate magnons are (meta)stable, we examine the existence of stable minima in the nonlinear potential energy $\mathcal{V}^{(4)}(\delta,\Phi)$. First, we compute the extrema of the nonlinear potential energy $\mathcal{V}^{(4)}(\delta,\Phi)$. We find the following critical points: 
\begin{align}
&\begin{array}{l}
\text{(i)} \quad \ds{1}=0, \quad \Phi_1 = 0 \\[0.2cm]
\text{(ii)} \quad \ds{2}=0, \quad \Phi_2 = \pi \\[0.2cm]
\text{(iii)} \quad \ds{3}=0, \quad \Phi_3 = \pm\arccos\left(\frac{\mathcal{C}_Q}{2\mathcal{D}_Q}\right) \\[0.2cm]
\text{(iv)} \quad \ds{4}=1-\left[ \dfrac{\mathcal{C}_Q \cos(\Phi_4)}
{\mathcal{A}_Q - \frac{\mathcal{B}_Q}{2} - \mathcal{D}_Q \cos\left(2\Phi_4\right) } \right]^{2}, \\ \phantom{\text{(iv)} \quad} \Phi_4 = 0 \\[0.2cm]
\text{(v)} \quad \ds{5}=1-\left[ \dfrac{\mathcal{C}_Q \cos(\Phi_5)}
{\mathcal{A}_Q - \frac{\mathcal{B}_Q}{2} - \mathcal{D}_Q \cos\left(2\Phi_5\right) } \right]^{2}, \\ \phantom{\text{(v)} \quad} \Phi_5 = \pi \\[0.2cm]
\text{(vi)} \quad \ds{6}=1-\left[ \dfrac{\mathcal{C}_Q \cos(\Phi_6)}
{\mathcal{A}_Q - \frac{\mathcal{B}_Q}{2} - \mathcal{D}_Q \cos\left(2\Phi_6\right) } \right]^{2}, \\[0.2cm]
\phantom{\text{(vi)} \quad} \Phi_6 = \pm \arccos\left(\frac{\mathcal{C}_Q}{2\mathcal{D}_Q}\right),
\end{array}
\label{eq:extrema}
\end{align}
where critical points (iv)-(vi) are extrema only if $\mathcal{C}_Q<0$.

To determine whether the nonlinear potential energy of the condensate magnons, given by Eq. (\ref{eq:v4u}), exhibits a minimum at any of these extrema and thus supports a (meta)stable state, we examine the sign of its second derivative. We define the discriminant as 
\begin{equation}
\text{Disc} = \biggl[
\biggl( \frac{\partial^2 \mathcal{V}^{(4)}}{\partial \delta ^2} \biggr)
\biggl( \frac{\partial^2 \mathcal{V}^{(4)}}{\partial \Phi ^2} \biggr)
-
\biggl( \frac{\partial^2 \mathcal{V}^{(4)}}{\partial \delta \partial \Phi} \biggr)^2
\biggr].
\label{eq:disc}
\end{equation}
The nonlinear potential energy has a minimum only if 
$\text{Disc} > 0$,
$( \frac{\partial^2 \mathcal{V}^{(4)}}{\partial \delta ^2} ) >0$, and
$( \frac{\partial^2 \mathcal{V}^{(4)}}{\partial \Phi ^2} ) >0$. 
These requirements depend on the numerical value of the interaction parameters
$\mathcal{A}_Q$, $\mathcal{B}_Q$, $\mathcal{C}_Q$ and $\mathcal{D}_Q$, which are functions of spin Hamiltonian parameters. 
We inspect the potential energy $\mathcal{V}^{(4)}(\delta,\Phi)$ and its second derivatives for our parameter space; see Fig. \ref{fig:graphics/part_der_i_iv.pdf}. We find that the critical point (ii) is the only minimum among all extrema, listed in \cref{eq:extrema}. 

\begin{figure}
\centering
\includegraphics[width=0.5\textwidth]{graphics/S_i_iv.pdf}
\caption{$\frac{\partial^2 \mathcal{V}^{(4)}}{\partial \delta ^2}$ (solid lines) and $\frac{\partial^2 \mathcal{V}^{(4)}}{\partial \Phi ^2}$ (dashed lines), as a function of easy-axis magnetic anisotropy strength $K_z/J$, for extrema (i), (ii), (iv), and (v). We set $\omega_D / \omega_\text{ex} = 0.3$ and $S=5/2$.}
\label{fig:graphics/part_der_i_iv.pdf}
\end{figure}

In \cref{fig:graphics/part_der_i_iv.pdf}, we plot $( \frac{\partial^2 \mathcal{V}^{(4)}}{\partial \delta ^2} )$ and $( \frac{\partial^2 \mathcal{V}^{(4)}}{\partial \Phi ^2} )$, as a function of easy-axis anisotropy strength $K_z/J$, for extrema (i), (ii), (iv), and (v). Within our model Hamiltonian, we find $|{\mathcal{C}_Q}/({2\mathcal{D}_Q})|>1$ and thus extrema (iii) and (iv) are not acceptable solutions. Note that $(\frac{\partial^2 \mathcal{V}^{(4)}}{\partial \delta \partial \Phi})=0$ for all extrema. We also find that \( C_Q > 0 \) for the entire range of parameter values in our model spin Hamiltonian, which renders extrema (iv)–(vi) unacceptable as solutions. Consequently, we conclude that only the critical point (ii) corresponds to a minimum in the effective condensation potential, Eq. (\ref{eq:v4u}), making a (meta)stable magnon condensate state possible.

\begin{figure}
\centering
\includegraphics[width=0.5\textwidth]{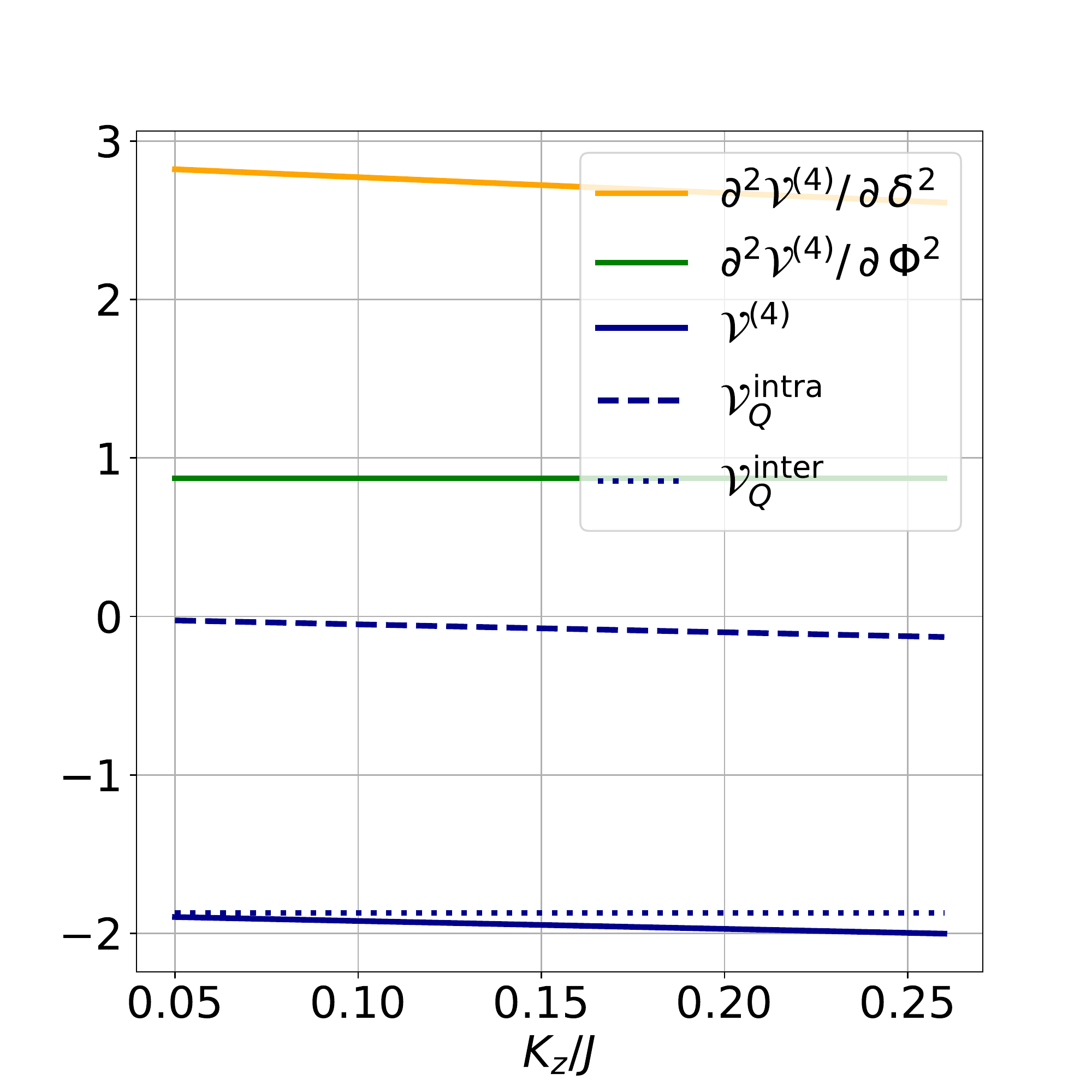}
\caption{
Total nonlinear potential energy of condensate magnons $\mathcal{V}^{(4)}$ (blue solid line) as a function of uniaxial magnetic easy-axis anisotropy, evaluated at its minimum, i.e., $\delta_2=0$ and $\Phi_2=\pi$. 
This potential energy can be separated into the intraband potential energy
$\mathcal{V}_Q^\text{intra}$ (blue dashed) and the interband potential energy 
$\mathcal{V}_Q^\text{inter}$ (blue dotted). 
We set $\omega_D / \omega_\text{ex} = 0.3$ and $S=5/2$.  }
\label{fig:pkt2}
\end{figure}

In \cref{fig:pkt2}, we plot the nonlinear magnon condensate potential energy $\mathcal{V}^{(4)}$ as a function of the easy-axis magnetic anisotropy strength $K_z/J$ for the extrema point (ii). 
We also plot the second derivative of the nonlinear potential energy with respect to the distribution difference $\delta$ and the phase sum $\Phi$. 
The plot shows the intra- and interband contributions to the potential energy. We find that the interactions are dominated by the interband potential energy. 
The state has a symmetric magnon distribution between the two magnon minima.
In FM systems with dipolar interactions, it has been shown that the condensate magnon state can be stabilized for both asymmetric and symmetric populations, depending on parameters such as anisotropy strength, thickness, and external field strength \cite{frostad2024anisotropy}.

\section{Zero-sound-like excitations in magnon BEC} \label{sec:zerosound}
The relative phase and relative density of a two-component condensate in a macroscopic BEC can be effectively described as canonical dynamical variables.
In the two-component magnon condensate, the difference between the two condensate phases can be a new Goldstone mode \cite{Li2013,Sun_2017}.
This out-of-phase mode in the collisionless regime shows similarities to the zero-sound phenomenon in Fermi liquid theory. 
In FM systems with dipolar interactions, two degenerate magnon condensates form at finite momenta, which can lead to the emergence of zero-sound-like excitations \cite{Li2013}.
On the other hand, in AFM systems, there are naturally two chiral magnons, leading to two-component magnon condensation states. Therefore, such zero-sound-like excitations may also arise in AFM systems. 
The condensate wave function is $\Psi(\bm{r})= \Psi_{\alpha} \text{e}^{-i \bm{Q}\cdot\bm{r}} + \Psi_{\beta}\text{e}^{i \bm{Q}\cdot\bm{r}}$, where $\Psi_{{(\alpha)\beta}} = \sqrt{n_{(\alpha)\beta}} e^{i \phi_{(\alpha)\beta}}$ and $n_{(\alpha)\beta}=N_{(\alpha)\beta}/V$ denote the density of two condensates in a system with volume $V$ \cite{Li2013}. The total condensate density in real space becomes $n(\bm{r})=n_{\alpha} + n_{\beta} + 2 \sqrt{n_{\alpha} n_{\beta}} \cos{(2\bm{Q}\cdot\bm{r}+\phi_{\beta}-\phi_{\alpha})}$. From this expression for the density, it becomes evident that the phase difference between two condensate states manifests itself solely in the real-space condensate density, effectively shifting the interference pattern as a whole. Meanwhile, the sum of these two phases contributes to the nonlinear potential energy of magnon condensates, see Eq. (\ref{eq:v4u}). Therefore, the Goldstone excitations of condensate magnons in AFM systems resemble zero-sound collective excitations, which are characterized by the absence of energy propagation.

We calculate the zero-sound spectrum by solving the corresponding Gross-Pitaevskii equation, following a similar approach to that in Ref. \cite{Li2013}. We begin with the Hamiltonian operator from \cref{eq:h1}, which provides the interaction terms listed in \cref{eq:h4interact_a,eq:h4interact_b}. The total energy density is then given by 
\begin{align} 
\frac{E}{V} &= \int d{\bm{r}} \biggl[
\frac{\hbar^2}{2m V} \big(\vert \nabla \Psi_{\alpha} \vert^2 + \vert \nabla \Psi_{\beta} \vert^2 \big)
\nonumber \\
&+ \mathcal{A}_Q \big(\vert \nabla \Psi_{\alpha} \vert^4 + \vert \nabla \Psi_{\beta} \vert^4 \big)
\nonumber + \mathcal{B}_Q \big(\vert \nabla \Psi_{\alpha} \vert^2 + \vert \nabla \Psi_{\beta} \vert^2 \big)
\nonumber \\
&+ \mathcal{C}_Q \big(\Psi_{\beta} \Psi_{\alpha} + \Psi_{\beta}^\dag \Psi_{\alpha}^\dag \big) \big(\vert \Psi_{\alpha} \vert^2 + \vert \Psi_{\beta} \vert^2\big)
\nonumber \\
&+ \mathcal{D}_Q \big(\Psi_{\beta} \Psi_{\beta} \Psi_{\alpha} \Psi_{\alpha} 
+ \Psi_{\beta}^\dag \Psi_{\beta}^\dag \Psi_{\alpha}^\dag \Psi_{\alpha}^\dag \big)
\biggr].
\end{align}
Here, $\hbar$ is the reduced Planck constant and $m$ is the effective mass of magnons at the band minima. In our model, see Eq. (\ref{energy}), we get $\hbar^2/(2m)\equiv a^2(\omega^2_\text{ex}+\omega^2_{D})/2\sqrt{\omega_{ez}^2-(\omega^2_\text{ex}+\omega^2_{D})}$.
In the previous section, we showed that the condensed magnons are stabilized for a symmetric magnon distribution. We now consider small deviations from this symmetric state, where $N_{\alpha(\beta)} = N_c/2 - (+) \den$ and $\Phi=\pi+\delta\phi$, with $\delta\phi=\phi_{\beta}-\phi_{\alpha}$. 
Using the canonical commutation relation between the relative phase and relative magnon density $[\dep,\den]=i$, and applying the Heisenberg equation of motion, we derive the following coupled equations of motion: 
%
\begin{subequations}
\begin{align}
\hbar \frac{\partial \den}{\partial t}
&= \frac{\hbar^2}{2m} N_c \nabla^2 \dep,\\
\hbar \frac{\partial \dep}{\partial t}
&= -\frac{\hbar^2}{2m} \frac{1}{N_c} \nabla^2 \den 
+ (\mathcal{A}_Q - \mathcal{B}_Q - \mathcal{D}_Q + \mathcal{C}_Q) V \den.
\end{align}
\end{subequations}
These equations of motion describe coupling between relative phase $\delta \phi$ and relative  density $\delta n$ of two-component magnon condensate states, leading to a zero-sound-like collective Goldstone mode.
To find the zero-sound dispersion, we apply the following harmonic oscillator ansatz for the relative density and relative phase, respectively: 
\begin{subequations}
\begin{align}
\den = \den_0 e^{i\omega t} e^{-i\bm{k}\cdot\bm{r}},
\\
\dep  = \dep_0 e^{i\omega t} e^{-i\bm{k}\cdot\bm{r}}.
\end{align}
\end{subequations}
Finally, we get the following dispersion relation for the zero-sound-like excitations:
\begin{equation}
\omega^2 = \frac{\hbar^2}{(2m)^2}k^4 
+ \frac{N_c}{2m}\left(\mathcal{A}_Q - \mathcal{B}_Q - \mathcal{D}_Q + \mathcal{C}_Q\right)  k^2.
\label{eq:zs}
\end{equation}
The gapless dispersion of \cref{eq:zs} is similar to the dispersion of the zero-sound excitation of an FM system with dipolar interaction in a symmetric condensed state; see Ref. \cite{Li2013}. In an exchange-dominant uniaxial AFM system, the coefficient of the $k^2$ term is always negative.

\section{Magnon BEC in biaxial AFM case} \label{sec:biaxial}
Now we consider a biaxial AFM system without DM interaction \cite{machado2017spin}. 
We focus on magnons at the bottom of the magnon bands, which, in this case, occur at the $\Gamma$ symmetry point. Therefore, we present only the two AFM resonance (uniform) modes \cite{machado2017spin}, 
\begin{align}
\omega_{\alpha(\beta)}^2 & = 
\left(\omega_{ez} + \frac{\omega_x}{2}\right)^2-\left(\frac{\omega^2_x}{4} + \omega_\text{ex}^2\right) + \omega_H^2 \nonumber \\
&- (+) 2\sqrt{
 \omega_H^2 \left[\left(\omega_{ez} + \frac{\omega_x}{2} \right)^2- \omega_\text{ex}^2 \right] +
 \frac{\omega_\text{ex}^2 \omega^2_x}{4}
}.
\end{align}
We see that in this case the energy degeneracy of two chiral magnon bands is broken even in the absence of the magnetic field, 
$\omega_H=0$, due to the presence of finite hard-axis magnetic anisotropy $\omega_x = S K_x$. 
Analytical calculation for the nonlinear magnon Hamiltonian of biaxial AFM systems in the presence of a magnetic field is very complicated \cite{machado2017spin,shiranzaei2022thermal}. To pursue the calculations analytically, we assume $J \gg K_x, K_z, h$, which is the case for AFM insulators such as NiO \cite{machado2017spin,rezende2019introduction}.
In this case, the interband magnon interactions are suppressed because of large splitting of two chiral magnon bands at the $\Gamma$ symmetry point. Since our focus is on the lowest magnon band, we restrict our calculations to the intraband magnon interactions of the $\beta$-magnon band.  
To compute four-magnon interactions at the lowest magnon band, i.e., $\beta$-magnon band, we use the following Bogoliubov transformation for uniform modes: 
\begin{subequations}
\begin{align}
\op{a} &= u_\alpha \op{\alpha} - v_\beta \bd{} ,\\
\op{b}^\dag &= -v_\alpha \op{\alpha} + u_\beta \bd{},
\end{align}
\end{subequations}
where the Bogoliubov coefficients at the $\Gamma$ symmetry point, within our approximations for strength of spin interactions, are given by \cite{machado2017spin}
\begin{subequations}
\begin{align}
u_{\alpha,\beta} &= \sqrt{
\frac{\omega_{ez} + \omega_x/2
+\omega_{\alpha,\beta}
}{
2 \omega_{\alpha,\beta}|_{\omega_H = 0}
}},\\
v_{\alpha,\beta} &= \sqrt{
\frac{\omega_{ez} + \omega_x/2 
-\omega_{\alpha,\beta}
}{
2 \omega_{\alpha,\beta}|_{\omega_H = 0}
}}.
\end{align}
\end{subequations}
The nonlinear intraband magnon interactions within the $\beta$-magnon band, in the uniform mode, can in general be expressed as
\begin{equation}
\mathcal{H}^{(4)}_\beta = 
\mathcal{A}_2 \big(\bo{} \bo{} \bd{} \bd{}\big) + \mathcal{A}_3 \big(\bo{} \bo{} \bo{} \bd{} + \text{H.c.}\big) 
+ \mathcal{A}_4 \big(\bo{} \bo{} \bo{} \bo{} + \text{H.c.}\big) 
\label{eq:biaxhp_v4o}
\end{equation}
To investigate the stability of condensed magnons, as before, we again perform Madelung's transformation, 
$ \bo{}  \rightarrow \sqrt{N_\beta} e^{i \phi_\beta}$,
where $\phi_\beta$ is the phase and $N_\beta$ is the number of condensed magnons.
Finally, we obtain a general expression for the nonlinear potential energy of condensed $\beta$-magnon, 
\begin{align}
\mathcal{V}^{(4)}_\beta = N_\beta^2 \left[
\mathcal{A}_2 + 2\mathcal{A}_3 \cos(2\phi_\beta) 
+ 2 \mathcal{A}_4 \cos(4\phi_\beta) \right],
\label{eq:biaxhp_v4}
\end{align}
where the interaction parameters are given by
\begin{subequations}
\begin{align}
\mathcal{A}_2  =& \frac{4}{N} \big[
-2J(u_\beta^2 v_\beta^2 ) 
+ J u_\beta v_\beta (u_\beta^2 + v_\beta^2) \nonumber \\
&- \frac{1}{8}K_z (u_\beta^4 + v_\beta^4) 
- 4 K_x(u_\beta^4 + v_\beta^4)\big], \\
\mathcal{A}_3  =& -\frac{8}{N}K_x(u_\beta^4 + v_\beta^4) ,\\
\mathcal{A}_4  =& 0.
\end{align}
\end{subequations}
We note that $\mathcal{A}_3$ appears purely due to the presence of the hard-axis magnetic anisotropy. In the biaxial case, unlike the uniaxial easy-axis configuration, the magnetic field influences the nonlinear interactions explicitly through the Bogoliubov coefficients.

The expression in \cref{eq:biaxhp_v4} shows that the nonlinear potential energy
$\mathcal{V}^{(4)}_\beta$ is proportional to $N_\beta^2$. This means that the expression for $\mathcal{V}^{(4)}_\beta$ does not have a nontrivial minimum for some finite number of condensed magnons. This indicates that there is no (meta)stable magnon condensation state for such systems. We conclude that a single-component magnon condensate state cannot be stabilized in magnetic systems. 
It is worth noting that, although we consider the lowest-energy magnons at the $\Gamma$ symmetry point, which are coherent, in the biaxial case, they still do not form a (meta)stable mBEC. This shows that although coherence is a hallmark of magnon condensates, not all coherent magnons form a condensate.

\section{Summary and Concluding Remarks} \label{sec:summary}
We have calculated the condensate interactions in a uniaxial and biaxial AFM system using both the \hp and \dm bosonization technique. The uniaxial easy-axis system in the absence of an external magnetic field, with a Rashba-like DM interaction, can host two energy-degenerate magnon condensate states at finite momenta, resembling the condensation observed in FM thin-film YIG. The nonlinear potential energy is a function of the phase sum $\Phi$ and the difference in the magnon number $\delta$. 
We analyzed the interactions for a range of easy-axis magnetic anisotropy strengths and DM interactions. We find that only the symmetric condensed magnon distribution $\delta=0$ leads to a stable magnon condensation. 
This indicates that even in the absence of dipolar interactions, which play a crucial role in the stability of two-component magnon condensates in FM thin-film YIG, it is possible to have a metastable magnon condensate state in AFM systems. 
We predict that the stable condensate state will have a symmetric distribution of magnons between the two condensate populations. Furthermore, we found the existence of a zero-sound-like Goldstone mode, which arises from the coupling between the oscillation of the relative phase $\delta \phi$ and the relative density $\delta n$ of the two-component magnon condensate state.
Even though the magnetic field along the easy axis does not explicitly appear in the magnon interaction terms, its application breaks the degeneracy of the two band minima, rendering the mBEC ground state unstable.
We propose that our theory can be tested experimentally in a uniaxial AFM material with Rashba-type DM interaction. Without an external magnetic field, this system has two chiral magnon band minima with the same energy at finite wave vector.
By applying a magnetic field, one can lift this energy degeneracy.
Our theory shows that a metastable two-component magnon condensate can be created in the absence of the magnetic field and that applying an external magnetic field along the magnetic ground state can destabilize it because of a reduction in the interband condensed magnon interactions.

In a biaxial AFM system, the degeneracy of the magnon energy minima is lifted even without an external magnetic field, resulting in only one chiral magnon population at the energy minimum. Therefore, due to the absence of interband magnon interactions, there is no (meta)stable condensate state. In reality, there can still be weak interband magnon interactions between two chiral magnons in the lowest $\beta$-magnon band and magnons at the minimum of the higher $\alpha$-magnon band, but we expect that such interactions would be weak to stabilize the condensation. It has been shown that applying a magnetic field perpendicular to both the easy- and hard-axis anisotropies can tune the magnon band splitting, leading to degeneracy at a critical field \cite{PhysRevB.107.184404}. In this case, we again expect a metastable two-component mBEC to persist.

\section*{Acknowledgement}
This project has been supported by the Research Council of Norway through its Centers of Excellence funding scheme, Project No. 262633 ``QuSpin''.

\appendix
\section{Uniaxial AFM case: \dm transformation} \label{sec:appendix}
We want to verify whether our results in the main text depend on the bosonization technique. We introduce the alternative \dm approach. Then, we follow the same procedure as in the \hp framework.
The \dm transformation gives us similar results for the noninteracting quadratic Hamiltonian of \cref{eq:h2}. When analyzing the four-magnon quartic Hamiltonian of \cref{eq:h4interact_a,eq:h4interact_b}, we find the same prefactors for 
$\mathcal{A}_Q$, $\mathcal{B}_Q$ and $\mathcal{D}_Q$. However, the prefactor $\mathcal{C}_Q$ must be modified. The new Hamiltonian terms in the \dm bosonization are given by, 
\begin{subequations}
\begin{align}
\mathcal{H}_Q^\text{intra } &=
\mathcal{A}_Q \bigl( \aq{} \aq{} \aqd{} \aqd{} + \bq{} \bq{} \bqd{} \bqd{}  \bigr)
,\\
\mathcal{H}_Q^\text{inter } &=
\mathcal{B}_Q \bigl( \aq{} \aqd{} \bq{} \bqd{}   \bigr) \nonumber \\
&+\mathcal{C}_Q^{+} \bigl( \aq{} \bq{}  \bq{}  \bqd{} +\aq{} \aqd{} \aqd{} \bqd{} \bigr) \nonumber \\
&+\mathcal{C}_Q^{-} \bigl( \aq{} \aq{} \aqd{} \bq{} + \aqd{} \bq{} \bqd{} \bqd{}  \bigr) \nonumber \\
&+\mathcal{D}_Q \bigl( \aq{} \aq{} \bq{} \bq{} + \aqd{} \aqd{} \bqd{} \bqd{}  \bigr). 
\end{align}
\end{subequations}
Here, we have defined
$\mathcal{C}_Q^{\pm} = \mathcal{C}_Q \pm \Delta_{\mathcal{C}}$ and $\Delta_C = -2 N^{-1}(3\eta_2 - \eta_1)[f_J(Q) + f_D(Q)]$. 
The prefactors $\mathcal{A}_Q$, $\mathcal{B}_Q$, $\mathcal{C}_Q$, and $\mathcal{D}_Q$ are listed in \cref{eq:prefacs}. 

Performing Madelung's transform, we obtain the following imaginary term that must be added to the  nonlinear potential energy of the condensed magnons in \cref{eq:v4u},
\begin{equation}
\mathcal{V}^{(4)}_C = \frac{N_c^2}{2}\sqrt{1-\ds{~}}
\left(i\frac{2\delta}{N_c} \sin(\Phi) \Delta_{\mathcal{C}}\right).
\label{eq:v4dm}
\end{equation}
Although the \dm transformation leads to different nonlinear Hamiltonian and nonlinear potential energy for condensed magnons, we note that the stable condensation state in our model is found at $\Phi=\pi$. Hence, this imaginary term vanishes and has no effect on the total potential energy of physical condensed magnons.

\bibliography{Refs}

\end{document}